\documentclass[draft]{agujournal2019}
\usepackage{url} 
\usepackage{lineno}
\usepackage[inline]{trackchanges} 
\usepackage{soul}
\usepackage{amsmath}
\draftfalse

\journalname{JGR: Space Physics}

\begin{document}

\title{Observations of Short-Period Ion-Scale Current Sheet Flapping}

\authors{L.~Richard\affil{1,2},
        	Yu.~V.~Khotyaintsev\affil{1},
        	D.~B.~Graham\affil{1},
        	M.~I.~Sitnov\affil{3},
        	O.~Le~Contel\affil{4},
        	P.-A. Lindqvist\affil{5}}

\affiliation{1}{Swedish Institute of Space Physics, Uppsala, Sweden}
\affiliation{2}{Space and Plasma Physics, Department of Physics and Astronomy, Uppsala University, Sweden}
\affiliation{3}{The Johns Hopkins University Applied Physics Laboratory, Laurel, Maryland, USA}
\affiliation{4}{Laboratoire de Physique des Plasmas, CNRS/Ecole Polytechnique IP Paris/Sorbonne Université/Université Paris Saclay/Observatoire de Paris, Paris, France}
\affiliation{5}{Space and Plasma Physics, School of Electrical Engineering, KTH Royal Institute of Technology, Stockholm, Sweden}

\correspondingauthor{L. Richard}{louis.richard@irfu.se}

\begin{keypoints}
	\item Kink-like flapping ion-scale current sheet (CS) propagating along the current direction is observed in the dusk-side plasma sheet.

	\item The wavenumber $k$ of the flapping oscillations scale with CS thickness $h$ as $kh \sim 1.15$, consistent with a drift-kink instability.

	\item Lower hybrid drift waves are observed when the CS is thinnest. Growth of the waves can be one of the factors limiting CS thinning.
 
\end{keypoints}

\begin{abstract}
 Kink-like flapping motions of current sheets are commonly observed in the magnetotail. Such oscillations have periods of a few minutes down to a few seconds and they propagate toward the flanks of the plasma sheet. Here, we report a short-period ($T\approx25$ s) flapping event of a thin current sheet observed by the Magnetospheric Multiscale (MMS) spacecraft in the dusk-side plasma sheet following a fast Earthward plasma flow. We characterize the flapping structure using the multi-spacecraft spatiotemporal derivative and timing methods, and we find that the wave-like structure is propagating along the average current direction with a phase velocity comparable to the ion velocity. We show that the wavelength of the oscillating current sheet scales with its thickness as expected for a drift-kink mode. The decoupling of the ion bulk motion from the electron bulk motion suggests that the current sheet is thin. We discuss the presence of the lower hybrid waves associated with gradients of density as a broadening process of the thin current sheet.
\end{abstract}

\newpage

\section{Introduction}
The kink-like flapping motion is a global oscillation of the magnetotail current sheet \cite{sergeev_orientation_2004,sergeev_survey_2006} and is correlated with fast flows \cite{sergeev_survey_2006}, which contributes to the explosive magnetotail activity \cite{sitnov_explosive_2019}. The period of these wave-like oscillations varies from couple of minutes \cite{zhang_wavy_2002,sergeev_current_2003,sergeev_orientation_2004,runov_electric_2005,zhang_neutral_2005,laitinen_global_2007,shen_magnetic_2008,wei_flapping_2015,rong_technique_2015,gao_distribution_2018,rong_cluster_2018,wang_ion_2019} to a couple of seconds \cite{wei_observations_2019} and they propagate with a phase velocity of the order of tens to hundreds of km~s$^{-1}$ in the spacecraft frame. Taking advantage of the abundance of the flapping events in the magnetotail, the statistical surveys carried out by \citeA{yushkov_current_2016} and \citeA{gao_distribution_2018} showed that the wave-like magnetotail current sheet observed in the flanks propagates away from the midnight sector. However, \citeA{gao_distribution_2018} also showed that around midnight sector ($|Y_{GSM}| < 6 R_E$) the flapping motion is dominated by up-down steady flapping motion.

The up-and-down motion in the midnight sector was proposed by \citeA{gao_distribution_2018} to be a mechanism of generation of the kink-like flapping of the current on both sides of the midnight sector. External disturbances such as Alfvénic fluctuations in the interplanetary magnetic field \cite{tsutomu_flapping_1976}, pressure pulses \cite{fruit_propagation_2002,fruit_propagation_2004} and solar wind directional changes \cite{wang_solar_2019} were proposed as possible mechanisms to trigger kink-like flapping motions. The magnetohydrodynamics eigenmodes analysis of the magnetic double gradient mechanism \cite{erkaev_magnetic_2008} pointed out that a gradient of the normal magnetic field induced by a local thinning of the current sheet, can trigger the flapping instability.

The kinetic treatment of the drift-kink and ion-ion kink instability proposed by \citeA{lapenta_kinetic_1997} and \citeA{karimabadi_ion-ion_2003} using a Harris model of the current sheet, concluded that the propagation of the wave-like current sheet was in the direction of the ion drift which agrees with direction of propagation of the flapping motions observed in the statistical studies \cite{yushkov_current_2016,gao_distribution_2018}. Indeed, \citeA{kissinger_diversion_2012} showed that the flows in the tail are diverging toward the flanks which coincided with the direction of propagation of the flapping motions. Using a model of an anisotropic thin current sheet, \citeA{zelenyi_low_2009} argued that kink-like flapping waves are low-frequency eigenmodes of the current sheet, which possess properties of drift waves. Indeed, observations of flapping events by Cluster reported by \citeA{zelenyi_low_2009} showed that kink-like flapping of thin current sheets propagate along the current density at velocities close to the relative drift velocity of ions and electrons.

The full particle simulation carried out by \citeA{sitnov_structure_2006} showed that the shape of the flapping wave of thin current sheets differs from the quasi-rectangular shape of thicker sheets. However, considering the normal electric field arising from the difference of velocities between ions and electrons at such scales, \citeA{sitnov_structure_2006} argued that the thinning of the current sheet is limited.

Nevertheless, there are several unsolved questions which are important for understanding how the flapping motions are generated, propagate and dissipate, which are in turn fundamental for understanding the tail structure, dynamics and energetics. These include: (1) What mechanism is responsible for generation of the kink-like flapping motions and how do they propagate? (2) How do ions behave in an ion scale flapping current sheet? (3) Does the thickness of the kink-like current sheet scale with its wavelength? (4) What processes balance the thinning of the current sheet? Here, we investigate the structure and dynamics of short period kink-like flapping of a thin ion-scale current sheet using multi-spacecraft observations by the Magnetospheric Multiscale (MMS) spacecraft \cite{burch_magnetospheric_2016} in order to advance our understanding of these questions.

\section{Observations}
We investigate a flapping event observed by the Magnetospheric Multiscale (MMS) mission \cite{burch_magnetospheric_2016} on 14 September 2019 between 07:54:00 UTC and 08:11:00 UTC. The flapping event is preceded by a sunward bursty bulk flow of ions and electrons associated with a dipolarization front. Unlike most of the previous reports of flappings, the event is not occuring in the proper magnetotail close to midnight. At this time, the MMS spacecraft were located on the dusk side of Earth ($X_{GSM}=-6.10~R_E$, $Y_{GSM}=12.93~R_E$, $Z_{GSM}=-0.14~R_E$). Among all of the instruments on board of the spacecraft, we use the magnetic field measured by the Flux Gate Magnetometer (FGM) \cite{russell_magnetospheric_2016}, the electric field measured by the Electric field Double Probe (EDP) \cite{ergun_axial_2016,lindqvist_spin-plane_2016}, the magnetic field fluctuations measured by Search Coil Magnetometer (SCM) \cite{lecontel_search-coil_2016}, and the particles moments measured by the Fast Plasma Investigation (FPI) \cite{pollock_fast_2016}. For all instruments we used the fast survey mode data except in section~\ref{ssec:lower-hybrid-drift-waves} where we used the burst mode EDP and SCM data. The spacecraft separation, $\sim 10$ km$\sim0.02d_i$, where $d_i=550$ km is the ion inertial length computed using the density averaged over the entire flapping event, is much smaller than the typical scale of the magnetotail current sheet. Hence, we present all quantities at the center of mass of the tetrahedron formed by the MMS constellation.

\subsection{Overview}
\begin{figure}[!b]
	\centering
	\includegraphics[width=\linewidth]{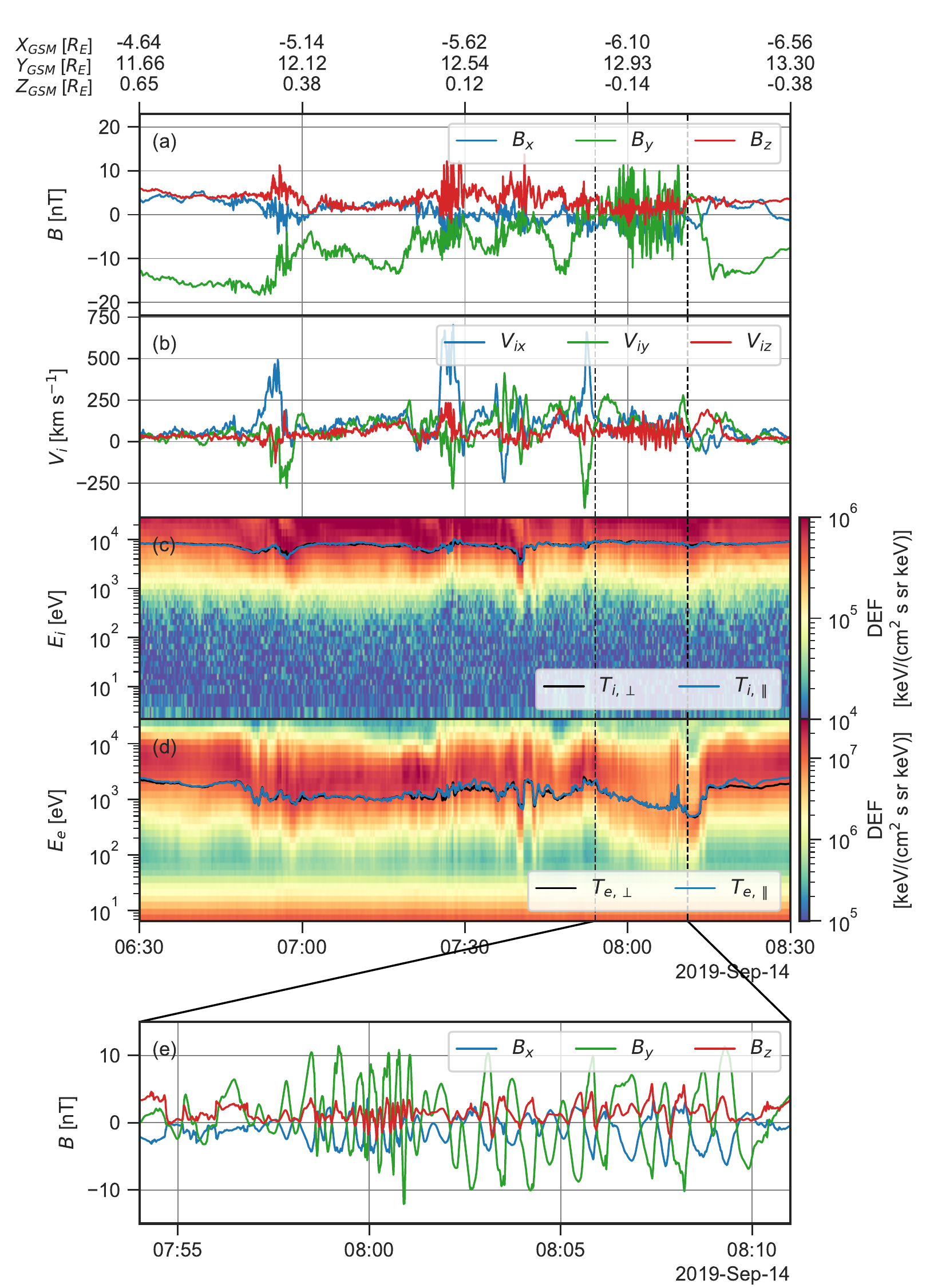}
	\caption{Overview of the flapping event observed by MMS on September 14th 2019. (a,e) Magnetic field in GSM coordinates, (b) ion bulk velocity in GSM coordinates, ion energy spectrum (c) and (d) electron energy spectrum.}
	\label{fig:overview}
\end{figure}

An overview of the event is shown in Figure~\ref{fig:overview}. We observe that the magnetic field is mainly along the Y axis (GSM) due to the duskside and northward location of MMS, which can be compared with the magnetotail configuration where the magnetic field is mainly in the X direction. The direction of the magnetic field is the same at the beginning and the end of the time interval, which indicates that the spacecraft remain on the same side of the plasma sheet. We observe a typical plasma sheet ion energy spectrum and ion temperature in Figure \ref{fig:overview}c. Both the magnetic field and the ion energy spectrum indicate that the spacecraft are far from the lobe and inside the plasma sheet. The ion bulk velocity, plotted in Figure~\ref{fig:overview}b, shows 3 jets. While we observe large-amplitude oscillations in the magnetic field during the two first jets, the third one is associated with a unidirectional northward magnetic field suggesting that it is a dipolarization front. We observe in Figure \ref{fig:overview}d a slow isotropic decrease in the electron temperature behind the dipolarization front. Following the dipolarization front embedded in the third plasma jet, we identify in Figure~\ref{fig:overview}e, large-amplitude oscillations of the three components of the magnetic field whose period varies from 10~s to 100~s, which is similar to the short-period flapping observed in Earth’s magnetotail by \citeA{wei_observations_2019}. Flapping motions with similar very short periods ($~13$s) were also reported in the Mercury magnetotail~\cite{zhang_flapping_2020}. We observe that $B_x$ and $B_y$ are anti-correlated during the flapping interval with a Pearson correlation coefficient $\rho = -0.84$. This anti-correlation is attributed to the flaring effect near the dusk flank \cite{fairfield_average_1979}. The observation of flapping motions in the wake of the dipolarization jet is consistent with the correlation between flapping motions and bursty bulk flows observed by \citeA{sergeev_survey_2006}.

\subsection{Ion Demagnetization and Ohm's Law}
\label{ssec:ion-demagnetization}
We need to find a coordinate system associated with the undisturbed current sheet. We apply the Minimum Variance Analysis to the magnetic field (MVAB) \cite{sonnerup_minimum_1998} over the entire time interval of the flapping motion ([07:54:00, 08:11:00]). The MVAB yields a large maximum to intermediate eigenvalue ratio, $\lambda_{max} / \lambda_{int} = 16.8$, meaning that the direction of maximum variance of the magnetic field $\mathbf{L} = [0.32, -0.95, -0.]$ GSM is well defined. The maximum variance directions obtained from the MVAB of the individual current sheet crossing are close to $\mathbf{L}$ (the median angle between the individual maximum variance directions and $L$ is 8.7 deg.). The L direction corresponds to the direction of the magnetic field i.e $\mathbf{B} \sim B_L \mathbf{L}$. On the other hand, we find a low value of the intermediate to minimum eigenvalue ratio, $\lambda_{int} / \lambda_{min} = 1.8$. This means that the minimum variance direction is not well defined, therefore we need to define the $\mathbf{M}$ and $\mathbf{N}$ directions in some other way. We choose $\mathbf{M}$ to point in the direction of the average ion bulk flow $\mathbf{M} \sim \langle \mathbf{V}_i \rangle / |\langle \mathbf{V}_i \rangle |= [0.62, 0.71, 0.34] $ GSM, since this is the expected direction of propagation of the flapping motion~\cite{kissinger_diversion_2012,yushkov_current_2016,gao_distribution_2018}. Then, the normal direction $\mathbf{N}$ is computed as $\mathbf{N} = \mathbf{L} \times \mathbf{M}/|\mathbf{L}\times \mathbf{M}| = [-0.36, -0.13, 0.92]$ GSM. Finally, the $\mathbf{M}$ direction is adjusted to make the LMN coordinates system orthogonal $\mathbf{M} = \mathbf{N} \times \mathbf{L} = [0.87, 0.3 , 0.38]$ GSM. We note that this direction is very close to the average current direction $ \langle \mathbf{J} \rangle / |\langle \mathbf{J} \rangle |= [0.94, 0.33, 0.11]$ GSM.

\begin{figure}[!b]
	\centering
	\includegraphics[width=\textwidth]{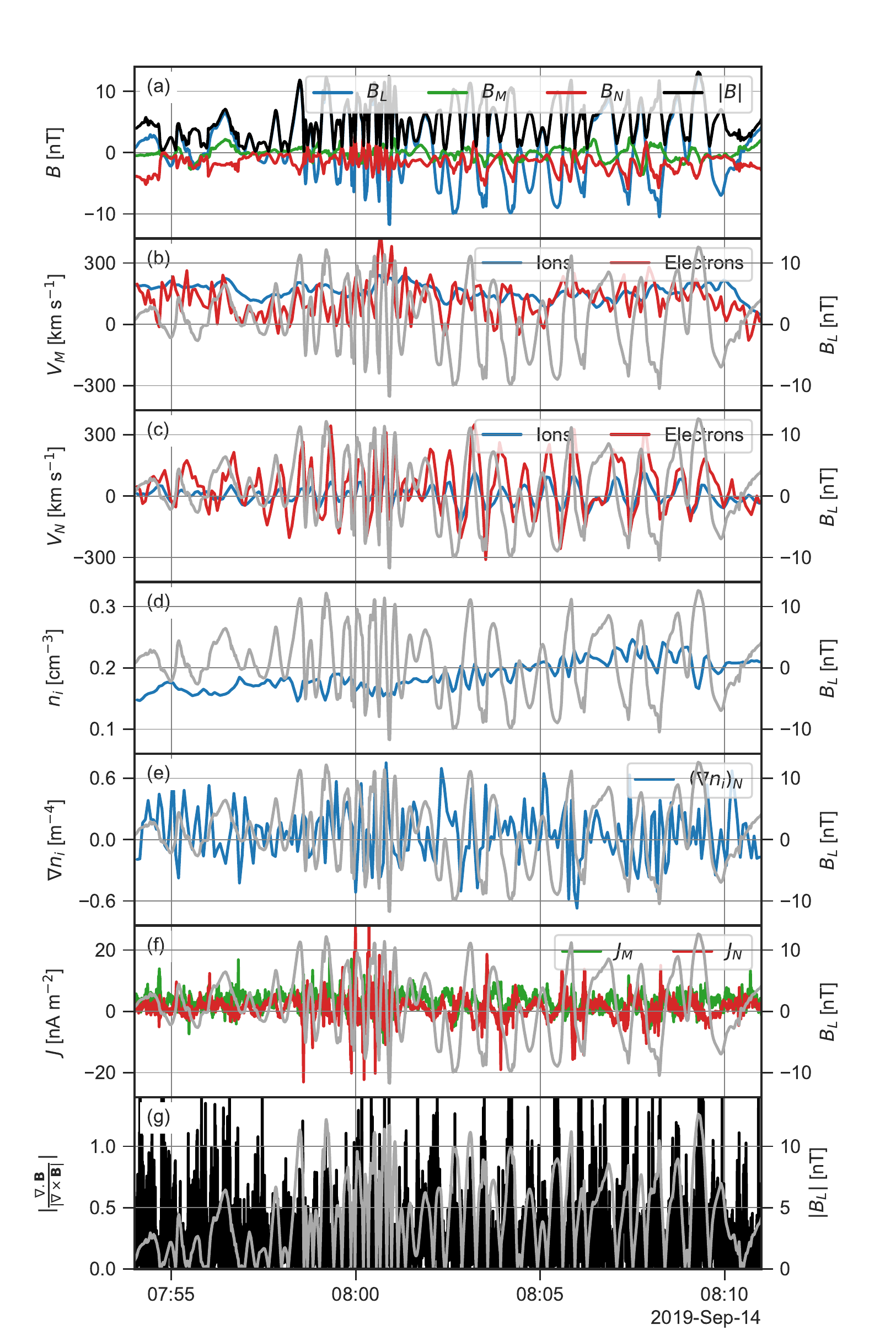}
	\caption{(a) Magnetic field in the global minimum variance frame LMN, (b) Ion bulk velocity (blue) and electron bulk velocity (red) in the $\mathbf{M}$ direction, (c) ion bulk velocity (blue) and electron bulk velocity (red) in the $\mathbf{N}$ direction, (d) ion number density, (e) ion density gradient in the $\mathbf{N}$ direction, (f) Current density from curlometer technique in the $\mathbf{M}$ (green) and $\mathbf{N}$ (red) directions. (g) Quality factor $\nabla . \mathbf{B} / |\nabla \times \mathbf{B}|$. Grey lines in panels (b-f) show the magnetic field $B_L$.}
	\label{fig:ion-demagnetization}
\end{figure}

First, we investigate the behavior of the ions during the oscillations of the magnetic field. From Figure~\ref{fig:ion-demagnetization}c we can see that the ion and electron bulk velocity normal to the mean current sheet plane, $V_{iN}$ and $V_{eN}$, are oscillating together with the magnetic field but have a phase shift of 90$^\circ$ with respect to $B_L$. In other words, the normal bulk velocity peaks in the center of the current sheet where $B_L = 0$, decreases as the spacecraft moves away from the current sheet, and reverses at the $|B_L|$ maxima. Hereafter, we call the region of maximum $|B_L|$ the current sheet boundary. The plasma density plotted in Figure \ref{fig:ion-demagnetization}d has a minimum at the current sheet boundaries and peaks in the current sheet center. This indicates the presence of density gradients normal to the current sheet, which is confirmed by estimate of the plasma density gradients using 4 spacecraft measurements shown in Figure~\ref{fig:ion-demagnetization}e. We observe that the gradients are larger during the short period oscillations ([07:59:30, 08:01:15]).

Figure~\ref{fig:ion-demagnetization}b shows that the M-component of the ion bulk velocity is rather constant during the time interval, $V_{iM}\sim 150$ km~s$^{-1}$, while $V_{eM}$ oscillates around the ion bulk velocity $V_{iM}$. Even though the ion and electron normal bulk velocities, $V_{iN}$ and $V_{eN}$, have the same phase, $V_{iN}$ is much lower than $V_{eN}$ so that the ion and electron motions are decoupled from each other. The large electron velocity results in a significant electron contribution to the current density.

Figure~\ref{fig:ion-demagnetization}f shows the current density ${\bf J}$ computed using the curlometer technique \cite{dunlop_analysis_1988}. The quality factor $\nabla . \mathbf{B} / |\nabla \times \mathbf{B}|$ plotted in Figure~\ref{fig:ion-demagnetization}g indicates that the current density is accurately computed, in particular, close to the current sheet ($|B_L| < 0$). $J_N$ oscillates around zero with the oscillation period similar to the magnetic field. This oscillating current density arises from the difference between the ion and electron velocities seen in Figures~\ref{fig:ion-demagnetization}b and \ref{fig:ion-demagnetization}c. The M-component of the current density $J_M$ oscillates with the same frequency as $J_N$ but with a 90$^\circ$ phase shift. We observe that $J_M$ remains positive during the flapping event, thus there is an overall current flowing in the $\mathbf{M}$ direction. $J_M$ peaks at the center of the current sheet and goes to zero at the current sheet boundaries where $V_{iM}$ and $V_{iN}$ reverse sign and the M-components of the flow are equal to each other, $V_{iM} = V_{eM}$.

\begin{linenomath}
To better characterise the ion and electron dynamics in the current sheet, we compute the different contributions to the Ohm's law,
\end{linenomath}

\begin{equation}
	\mathbf{E} + \mathbf{V}_i\times\mathbf{B}~=~ \frac{\mathbf{J}\times\mathbf{B}}{ne} - \frac{\nabla P_e}{ne},
\end{equation}
where $\mathbf{E}$ is the electric field measured by the EDP, $\mathbf{V}_i\times \mathbf{B}$ is the ion convection term, $\left (\mathbf{J}\times\mathbf{B}\right ) /ne$ is the Hall term and $\left (\nabla P_e\right ) /ne$ is the electron pressure gradient term.

\begin{figure}[!b]
	\centering
	\includegraphics[width=\textwidth]{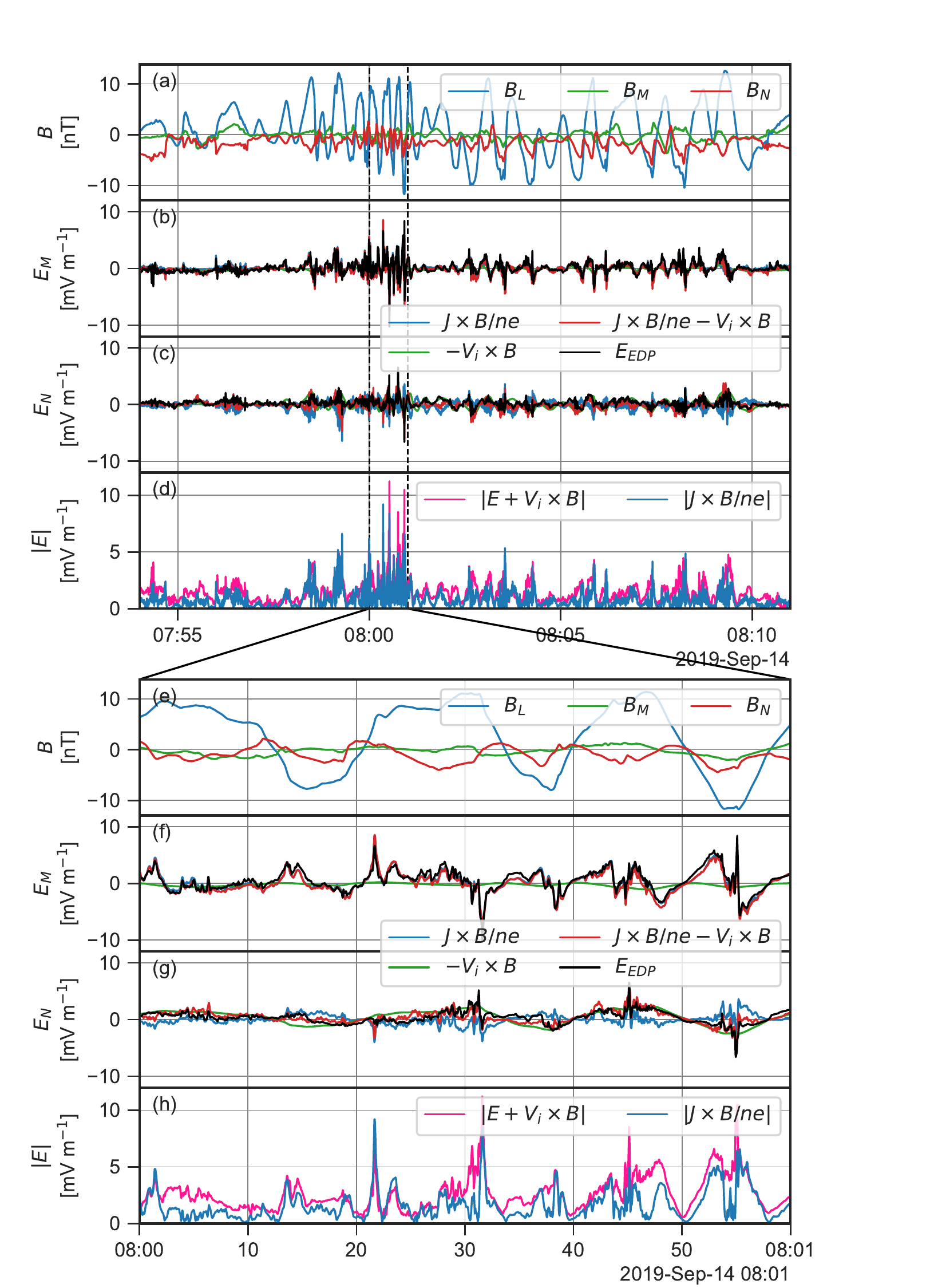}
	\caption{(a, e) Magnetic field in the global LMN coordinates system, (b, c, f, g) electric field perpendicular to the maximum variance magnetic field with Hall contribution (blue), ion convection (green), Hall + ion convection (red) and electric measured by the EDP (black). (d, h) Magnitude of the non-ideal electric field (pink) and the Hall contribution (blue)}
	\label{fig:ohms-law}
\end{figure}

Figure~\ref{fig:ohms-law} shows M and N components of these terms, as well as the non-ideal electric field $|\mathbf{E} + \mathbf{V}_i\times \mathbf{B}|$. We observe that $E_M$ is the largest electric component (Figures~\ref{fig:ohms-law}b and~\ref{fig:ohms-law}f). $E_N$ is balanced by the ion convection term, while the N component of the Hall term is small. $E_M$ is spiky with peaks reaching $E_M \approx 10$ mV~m$^{-1}$. The peaks are located away from the center of the current sheet ($B_L=0$). We see in Figures~\ref{fig:ohms-law}b and~\ref{fig:ohms-law}f that the electron convection $\mathbf{V}_e \times \mathbf{B} = \mathbf{J} \times \mathbf{B} / ne - \mathbf{V}_i \times \mathbf{B}$(red) is in good agreement with the measured electric field (black) $\mathbf{E_M} \approx - \left ( \mathbf{V}_e \times \mathbf{B} \right )_M$, which indicates that the electrons are magnetized. On the other hand, the ion convection term $\left ( \mathbf{V}_i \times \mathbf{B} \right )_M$ (green) does not balance the M-component of the electric field. We observe in Figures~\ref{fig:ohms-law}d and~\ref{fig:ohms-law}h, that the non-negligible non-ideal electric field in the MN plane $\left | \mathbf{E} + \mathbf{V}_i \times \mathbf{B} \right | $ (pink), is well balanced by the Hall term $\left | \mathbf{J} \times \mathbf{B} / ne\right |$ (blue). This means that the contribution from the electron pressure gradient term is small. The Hall electric field comes from the decoupling of the ion motion from the motion of electrons, and such decoupling is possible at scales below the ion inertial length. Therefore, the dominant contribution of the Hall field to the non-ideal electric field indicates that the electrons are magnetized while the ions are partially demagnetized, which indicates that we observe an ion-scale current sheet.

An illustration of the electron motions in the frame of the ions is shown in Figure \ref{fig:current-sheet}d, where the red dots indicate the electrons. The demagnetized ions move mainly in the ${\mathbf M}$ direction, while the electrons remain frozen-in and follow the oscillations of the current sheet in the N direction.

The Hall electric field can provide information on the degree of kinking of the current sheet. For a planar (undisturbed) current sheet the Hall electric field is expected to be normal to the current sheet, $E_N \sim J \times B / ne$. For a disturbed (kinked) current sheet, the current density has two components $J_N$ and $J_M$ in the MN plane, and therefore the Hall electric field will have two components, $E_M$ and $E_N$. We observe that $J_N$ changes sign at the current sheet boundary (Figure~\ref{fig:ion-demagnetization}f), and therefore $E_M$ should also change sign there, which is what we observe. On the other hand, $J_M$ remains always positive (Figure~\ref{fig:ion-demagnetization}f), but $B_L$ reverses at the current sheet center, so that $E_N$ should also reverse there, which is also consistent with the observations. For a strongly-kinked current sheet such that the current at the crossing of the current sheet center is mostly in the N-component (the current sheet is locally nearly vertical), we expect $E_M \sim J_{N} \times B_L / ne$ to be the dominant component, which is observed during the interval of short-period oscillations ([07:59:30, 08:01:15]). Therefore, the observed behavior of the Hall electric field indicates that the observed current sheet is kinked, and the kinking is stronger during the interval of short period oscillations.

\subsection{Current Sheet Width}
\label{ssec:current-sheet-width}
\begin{figure}[!b]
	\centering
	\includegraphics[width=\textwidth]{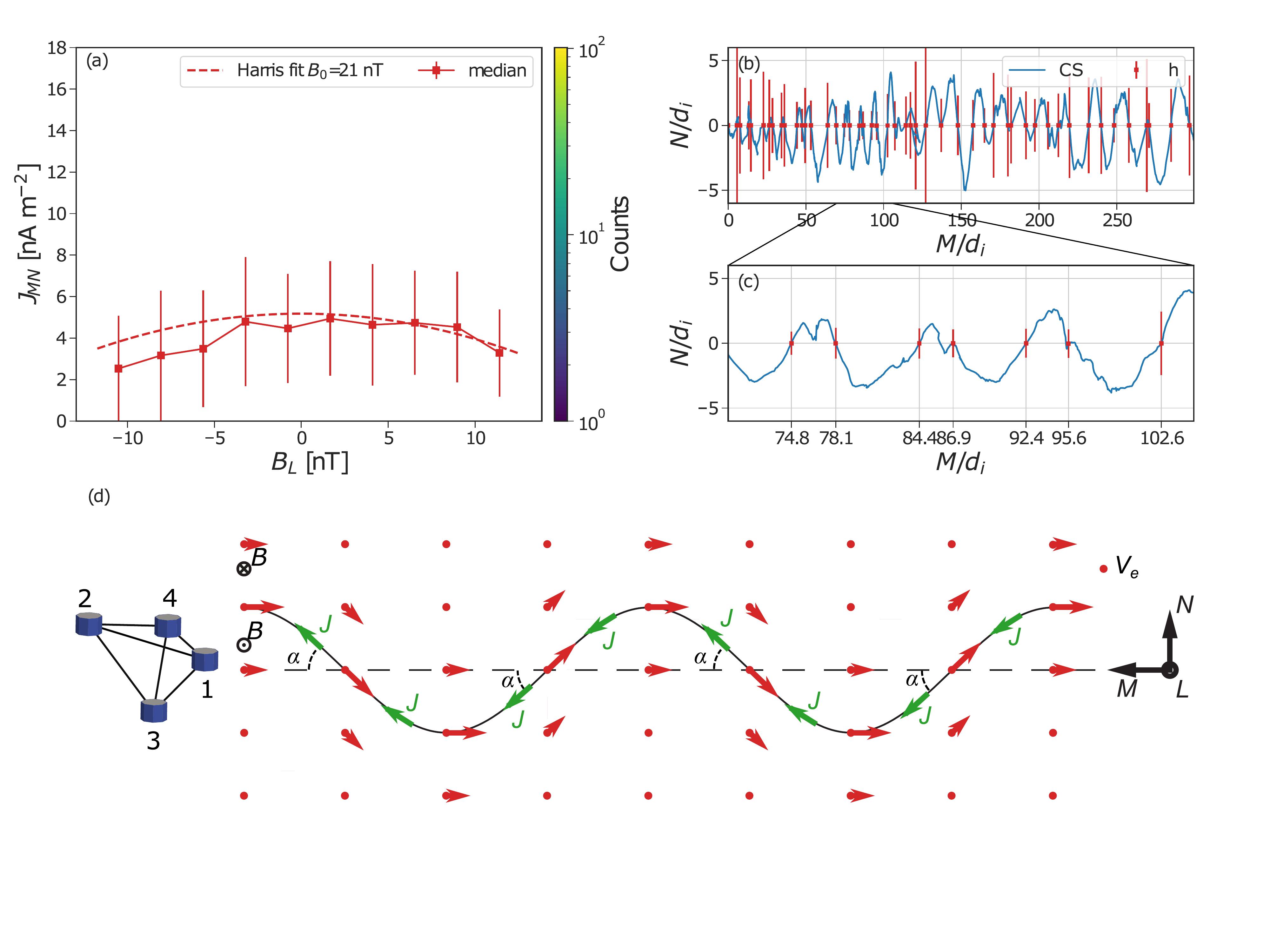}
	\caption{(a) Histogram of the current density with respect to the magnetic field. The red squares represent the median current density and the errorbars the corresponding standard deviation for 2 nT wide bins of the magnetic field, and the red solid line is the Harris approximation of the current density \ref{eq:harris}. (b, c) Two-dimensions geometry of the oscillating current sheet (blue solid line) and its thickness (red bars). (d) Interpretation of the electron and current sheet motion in the ion frame.}
	\label{fig:current-sheet}
\end{figure}

\begin{linenomath}
To fully characterize the geometry of the oscillating current sheet, we now focus on its thickness. First, we plot the distribution of the magnitude of the current density $J_{MN}$ with respect to the magnetic field $B_L$ in Figure \ref{fig:current-sheet}a as a 2D histogram. We compare this distribution to the scaling for the Harris model (red dashed line):
\end{linenomath}

\begin{equation}
	J_{MN} = \frac{B_0}{\mu_0 h_{H}} \left [ 1 - \left ( \frac{B_L}{B_0}\right )^2 \right ],
	\label{eq:harris}
\end{equation}
 where $B_0$ is the field outside of the current sheet (lobe field) and $h_{H}$ is the current sheet thickness. We see that the Harris scaling with $h_{H} = 6.0 \pm 0.2 d_i$ and $B_0 = 21 \pm 1$ nT fits the observed distribution rather well.
 
 The lobe field $B_0$ can be alternatively obtained from the pressure balance condition $B_{0PBC} = (B^2 + 2 \mu_0 P_i)^{1/2}$, which for our case yields $B_{0PBC} \sim 28$ nT. However, Figure~\ref{fig:current-sheet}a shows that the current density becomes very small for a much smaller value of $B_0 = 21$ nT, which indicates that the observed thin and intense current sheet is likely embedded into a thicker one. This situation resembles the current distributions Figure 5 in \citeA{runov_local_2006}.

\begin{linenomath}
The thickness $h_{H}$ obtained above is a global average current sheet thickness.
In order to access the local scale of the core current sheet, we compute the thickness of the current sheet at every crossing of the current sheet center. We estimate the local thickness $h$ of the current sheet as:
\end{linenomath}

\begin{equation}
	h = \frac{B_0}{\mu_0 |\mathbf{J}|_{B=0}},
\label{cswidth}
\end{equation}
where $|\mathbf{J}|_{B=0}$ is the current density estimated by the curlometer technique at the neutral line ($B_L=0$) and $B_0 = 21$ nT is the magnetic field at the current sheet boundaries obtained earlier. Here we use a constant lobe field $B_0$ which is a reasonable assumption since $B_{0PBC}$ does not vary significantly during the event $\sigma_{B_{0PBC}} / \langle B_{0PBC} \rangle = 4 \% $, where $\sigma_{.}$ denotes the standard deviation, and as the time interval of the flapping of motion is relatively short (17 minutes). Figures~\ref{fig:current-sheet}b and~\ref{fig:current-sheet}c show the thickness (red vertical lines) computed using equation (\ref{cswidth}) normalized by the ion inertial length $d_i=550$ km. Also $d_i$ is assumed to be constant because the plasma density ( Figure~\ref{fig:ion-demagnetization}d) does not change significantly during the interval ($\sigma_{n_{i}} / \langle n_{i} \rangle = 12 \%$), and the ion inertial length $d_i$ does not change significantly either ($\sigma_{d_{i}} / \langle d_{i} \rangle = 6 \%$). The abscissa and ordinate are the spatial scales computed using the spatio-temporal derivative technique (discussed below) in the direction of propagation of the wave-like structure. We see that the thickness of the current sheet experiences significant variations. The thinnest current sheet is observed during the interval 07:59:30--08:01:15 UT which we show in detail in Figure \ref{fig:current-sheet}c. This interval corresponds to the interval where the oscillations of $\mathbf{B}$ and plasma parameters have the smallest period (see Figure~\ref{fig:ion-demagnetization}). The thinnest current sheet we observe has $h \sim 1.8 d_i$, which is more than three times thinner than the average thickness $h_{H}$. Thus, at this time the current sheet is very thin, comparable to the ion inertial length scale, which is consistent with the ion demagnetization discussed above.

\begin{linenomath}
In order to estimate the local scale of the oscillating current sheet and its geometry, we compute the local velocity of the wave-like structure using the spatio-temporal derivative (STD) method \cite{shi_motion_2006}. The dimensionality of the structure has to be known before computing the velocity. Here, when the spacecraft are within the current sheet, the eigenvalues of the rotation rate tensor $\mathbf{S}$, are ordered as: $\lambda_{1} \gg \lambda_{2} \gg \lambda_{3}$. In such a case, as mentioned by \citeA{shi_dimensionality_2019}, the structure can be either 1D or 2D. However, from the MVAB analysis of the individual current sheet crossings we find $\lambda_{max} \gg \lambda_{int} \sim \lambda_{min}$, which indicates that the individual crossings are 1D structures. These individual 1D structures share the same maximum variance direction (along $\mathbf{B}$) and have different normal directions, and thus the flapping current sheet as a whole is two-dimensional. Assuming that the structure is quasi-stationary, we compute the velocity of the structure as
\end{linenomath}

\begin{equation}
	\label{eq:std}
	\mathbf{V}_{str} = -\textrm{d}_t \mathbf{B} \left [\nabla \mathbf{B}\right]^T \left [\mathbf{S}\right ]^{-1}
\end{equation}
where $\mathbf{V}_{str}$ is the velocity of the structure and $\mathbf{S}=\nabla \mathbf{B} \left ( \nabla \mathbf{B}\right )^T$ is the rotation rate tensor in the $\mathbf{LMN}$ coordinates system. Using the approach similar to \citeA{shi_motion_2006} we can reduce the system from three to two equations. Assuming that the structure is 2D, the gradients of the magnetic field are well defined in the perpendicular to the magnetic field plane, but not along the magnetic field. Thus, only the projection of the equation \ref{eq:std} onto a plane perpendicular to the magnetic field may give an accurate result. Therefore, we project equation~\ref{eq:std} onto the $MN$ plane. We compute the time derivative of the magnetic field using a second order accurate central differences with a step size taken to be 2.5 s ($\sim T_{min} / 4 $, where $T_{min}$ is the shortest period of oscillations of $B_L$ ). We discard non-physical values due to inversion of the rotation rate tensor $\mathbf{S}$ using the following criteria on the determinant of $\mathbf{S}$: $\det (\mathbf{S}) > \langle \det (\mathbf{S}) \rangle_T - \sigma^T_{\det (\mathbf{S})}$, where $\langle \det (\mathbf{S}) \rangle_T$ and $\sigma^T_{.}$ denote the average and the standard deviation over one period of oscillations of $B_L$.

\begin{figure}[!b]
	\centering
	\includegraphics[width=\textwidth]{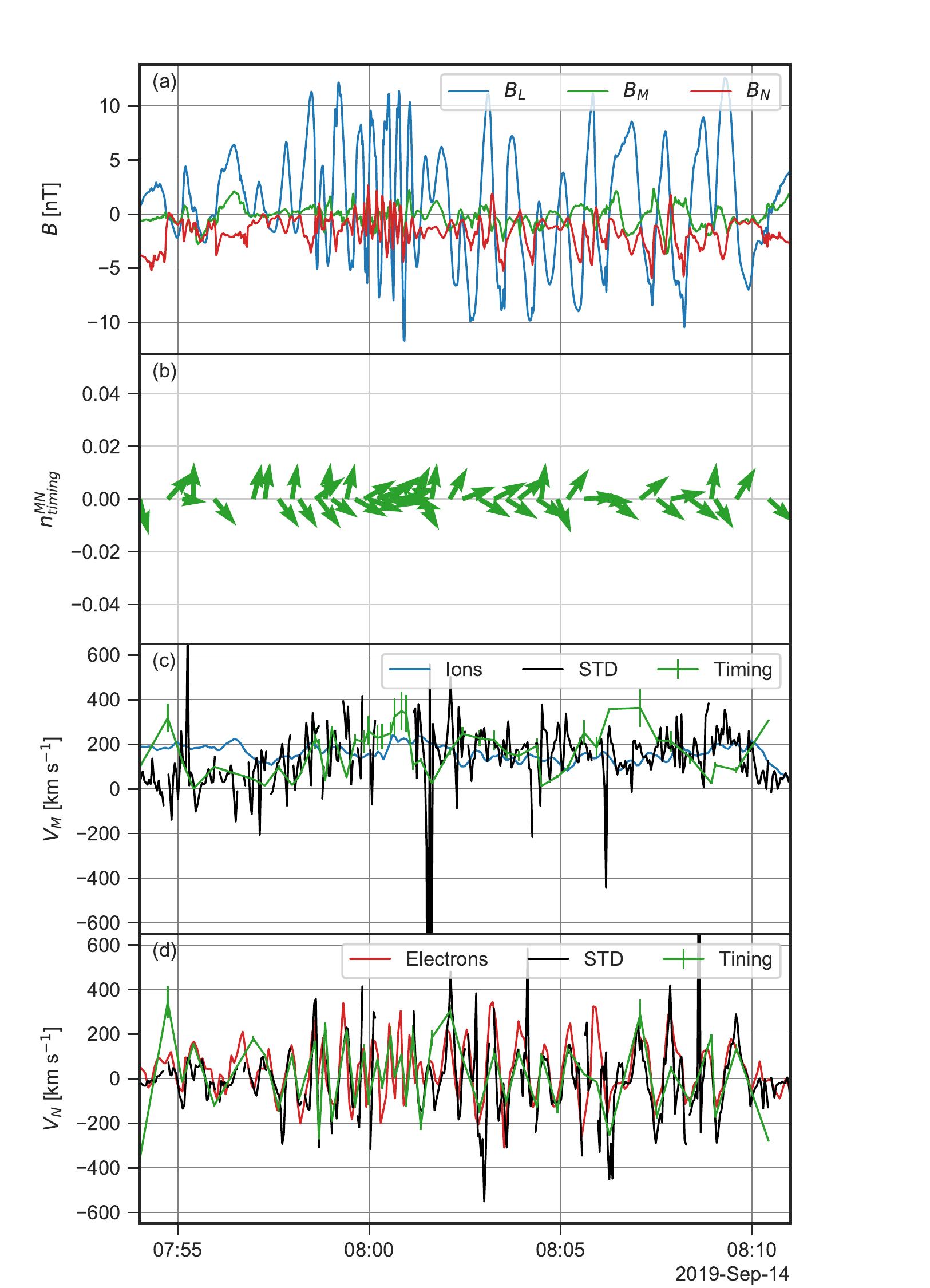}
	\caption{(a) Magnetic field in the global LMN coordinates system. (b) Ion velocity (blue) and velocity of the current sheet from spatiotemporal derivative (black) and timing (green) along the $\mathbf{M}$ direction. (c) Electron velocity (red) and velocity of the current sheet from spatiotemporal derivative (black) and timing (green) along the flapping direction.}
	\label{fig:structure-velocity}
\end{figure}

In addition to the STD method which provides a continuous velocity estimate, we use the timing method \cite{vogt_accuracy_2011} at every crossing of the current sheet ($B_L = 0$). The timing method gives the normal $N_{TM}$ to the structure and its associated velocity. We project the velocities from STD and timing onto the M (Figures \ref{fig:structure-velocity}c) and N (Figures \ref{fig:structure-velocity}d) directions. The two methods provide similar results, which agree particularly well during the short period train 07:59:30--08:01:15 UT, where the velocity increases. We also note that the obtained velocity of the flapping structure is close to the ion bulk velocity (blue). Thus we conclude that the flapping structure propagates in the $+\mathbf{M}$ direction (the average current direction) and the structure is approximately stationary in the ion frame of reference.

Figure~\ref{fig:structure-velocity}b shows the projection of the normal obtained using the timing method onto the $\mathbf{MN}$ plane where the abscissa represents the M direction and the ordinate represents the N direction. \citeA{rong_technique_2015} proposed a technique to qualitatively diagnose the flapping type, in particular they found that if the polarity of $\kappa = \operatorname{sign}{(n_M \times n_N)} \times \operatorname{sign}{(\Delta B_L)}$ is constant during multiple crossings, then the flapping motion is kink-like and it is propagating in the $\operatorname{sign}{(\kappa)}$ M direction. We observe that $n_M > 0$ while $n_N$ changes sign according to the polarity variation $\Delta B_L$ which corresponds to $\kappa = +1$ for every crossings of the current sheet. Hence, this confirms that the observed flapping is of a kink-like type and it propagates in the +M direction (the average current direction).

Integrating the velocity of the current sheet given by the STD method, $\mathbf{V}_{str}$, over time, we obtain the current sheet geometry in the $\mathbf{MN}$ plane. Assuming that the spacecraft are stationary while the oscillating current sheet moves past the spacecraft, we plot in Figures~\ref{fig:current-sheet}b and \ref{fig:current-sheet}c the position of the current sheet center with respect to the spacecraft (blue). We observe that as the period of the oscillations changes in time from 10 s to 1 min, the wavelength of the oscillating current sheet varies from 5 to $10~d_i$, but remains comparable to the peak-to-peak amplitude of the oscillations $\sim 4 d_i$.

The maximum value of the tilt angle $\alpha = \operatorname{tan}^{-1}\left.\left( J_{N} / J_{M}\right )\right|_{B=0}$, shown in Figure \ref{fig:current-sheet}d, between the current sheet and the $\mathbf{LM}$ plane is $\sim45^{\circ}$ which gives a wavy shape to the current sheet. The maximum value of the tilt is observed in the region where the thickness of the current sheet is the smallest. Similar correlation between the tilt angle and the thickness was also observed during multiple flapping events by Cluster \cite{shen_magnetic_2008,rong_analytic_2010}. The picture given by the Figure~\ref{fig:current-sheet}d is either compressed or stretched in the direction of propagation of the wave-like structure as the wavelength becomes shorter or longer. Hence, for a short wavelength the tilt angle $\alpha$ increases, while in the special case of an infinite wavelength we recover the 1D picture of a thin current sheet.

\subsection{Dispersion Relation}
\label{ssec:dispersion-relation}
\begin{figure}[!b]
	\centering
	\includegraphics[width=\textwidth]{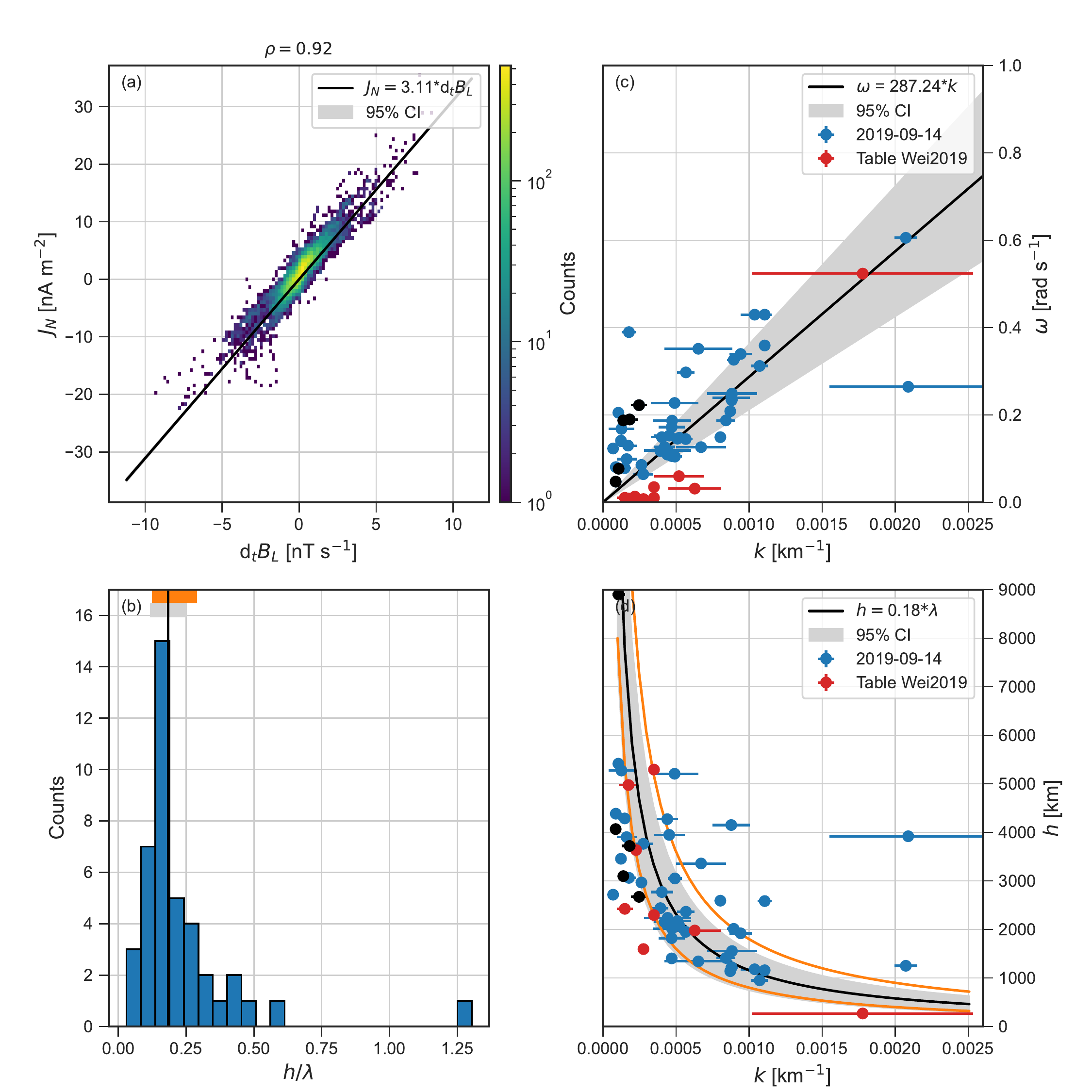}
	\caption{(a) Normal current density $J_N$ with respect to the time derivative of the magnetic field $\textrm{d}_t B_L$. The colors indicate the distribution of the values and the black line a $y=ax$ fit (see text). (b) Distribution of the thickness to wavelength ratio during the event. (c) Frequency $\omega$ of the oscillations of the magnetic field $B_L$ with respect to the wavenumber $k$ given by the timing method. (d) Current sheet thickness with respect to the wavenumber $k$ given by the timing method. The blue dots in panels (c) and (d) are the current sheet crossings of the event reported in this paper and the red dots are other studies of flapping motions in the tail summarised by \citeA{wei_observations_2019}. The black dots are the crossings far from our event that are considered as outliers. The orange lines in panel (d) and orange bow in panel (b) show the lower and upper bounds of the theoretical prediction of the range of wavenumber in terms of current sheet thickness which maximize the growth rate of the drift-kink instability.}
	\label{fig:disperel}
\end{figure}

We showed in section~\ref{ssec:current-sheet-width} that the wave-like structure is propagating along the current direction. We now focus on the estimation of the phase velocity of this wave-like motion. To do so, we use the velocity of the structure obtained by the timing method applied to the crossings of the current sheet center. We compute the wavelength using $\lambda = V_{\phi}T$, where $V_{\phi} = V_{TM} / \sin(\alpha)$ is the phase speed from the timing method with $V_{TM}$ being the velocity of the structure along the normal $n_{TM}$, $\alpha$ is the tilt angle of the current sheet, and $T$ is the period of the oscillations which can be estimated as the time between two current sheet boundaries. To plot the result in a more usual way, we plot in the Figure~\ref{fig:disperel}c the angular frequency $\omega = 2\pi/T$ versus the wavenumber $k = 2\pi/\lambda$. In order to compare with other studies, we use the data provided by~\citeA{wei_observations_2019} (red dots). The crossings of the current sheet which are not related to the main phase of flapping motion are considered as outliers and marked by the black dots. Finally, a linear fit is applied, which yields the phase speed $V_{ph}\approx 300 \pm 40~\textrm{km}~\textrm{s}^{-1}$, which is consistent with the very short period current sheet flapping observed by \citeA{wei_observations_2019}

We recall that the current density normal to the mean plane $J_N$ oscillates around zero with the same period as the magnetic field (Figure \ref{fig:ion-demagnetization}f). Figure~\ref{fig:disperel}a shows a 2D histogram of $J_N$ as a function of the time derivative of the magnetic field $\textrm{d}_t B_L$ where the color represents the counts. In Figure~\ref{fig:disperel}a we observe a strong correlation between $J_N$ and $\textrm{d}_t B_L$, with a Pearson correlation coefficient of $\rho = 0.93$. Hence, $\mu_0 J_N = a \mu_0 \textrm{d}_t B_L$ where $a$ is a coefficient of proportionality. Using the Ampere’s law:

\begin{eqnarray}
	\mu_0 J_N & = & \left (\nabla\times \mathbf{B}\right)_N\nonumber\\
     	&=& \left( ik_LB_M - ik_M B_L\right)\nonumber\\
     	&\approx& - i k_M B_L. \nonumber
\end{eqnarray}

\begin{linenomath}
Thus, the correlation between $J_N$ and $\textrm{d}_t B_L$ can be written as :
\end{linenomath}

\begin{equation*}
	-i k_M B_L \approx i\mu_0 a \omega B_L,
\end{equation*}

\begin{linenomath}
And then the phase velocity is
\end{linenomath}

\begin{equation}
	V_{ph} = \frac{\omega}{k_M} \approx -\left [a\mu_0\right]^{-1}.
\end{equation}
The linear fit $\mu_0J_N = a\mu_0\textrm{d}_t B_L$ gives $a = 3.12 \pm 0.01$ H$^{-1}$~s, which in turns gives $V_{ph} = 255 \pm 1$ km~s$^{-1}$ consistent with the timing method.

Finally, as shown in Figures~\ref{fig:current-sheet}b and \ref{fig:current-sheet}c, the thickness of the current sheet decreases as the wavelength decreases. Thus, we now investigate a possible scaling of the thickness of the current sheet with the wavelength. Figure~\ref{fig:disperel}d shows the thickness of the current sheet $h$ as a function of the wavenumber $k$. As in Figure~\ref{fig:disperel}c we also provide the values from \citeA{wei_observations_2019}. Figure~\ref{fig:disperel}b presents the distribution of the scales of the thickness with respect to the wavelength of the oscillating current sheet. Using a $1/k$ fit (black line) on the values reported here only, we conclude that the thickness scales with the wavelength as $h \approx (0.18 \pm 0.03) \lambda$.

The lower and upper bounds of the theoretical range of the wavenumber in terms of current sheet thickness obtained in the linear analysis of drift-kink instability \cite{zelenyi_low_2009} are shown by the orange lines in Figure \ref{fig:disperel}d (orange box in Figure \ref{fig:disperel}b). The theoretical prediction agrees with the observed scales which suggests that the drift-kink instability may be responsible for the observed wave-like motion of the current sheet.

\subsection{Lower hybrid drift waves}
\label{ssec:lower-hybrid-drift-waves}
\begin{figure}[!b]
	\centering
	\includegraphics[width=\textwidth]{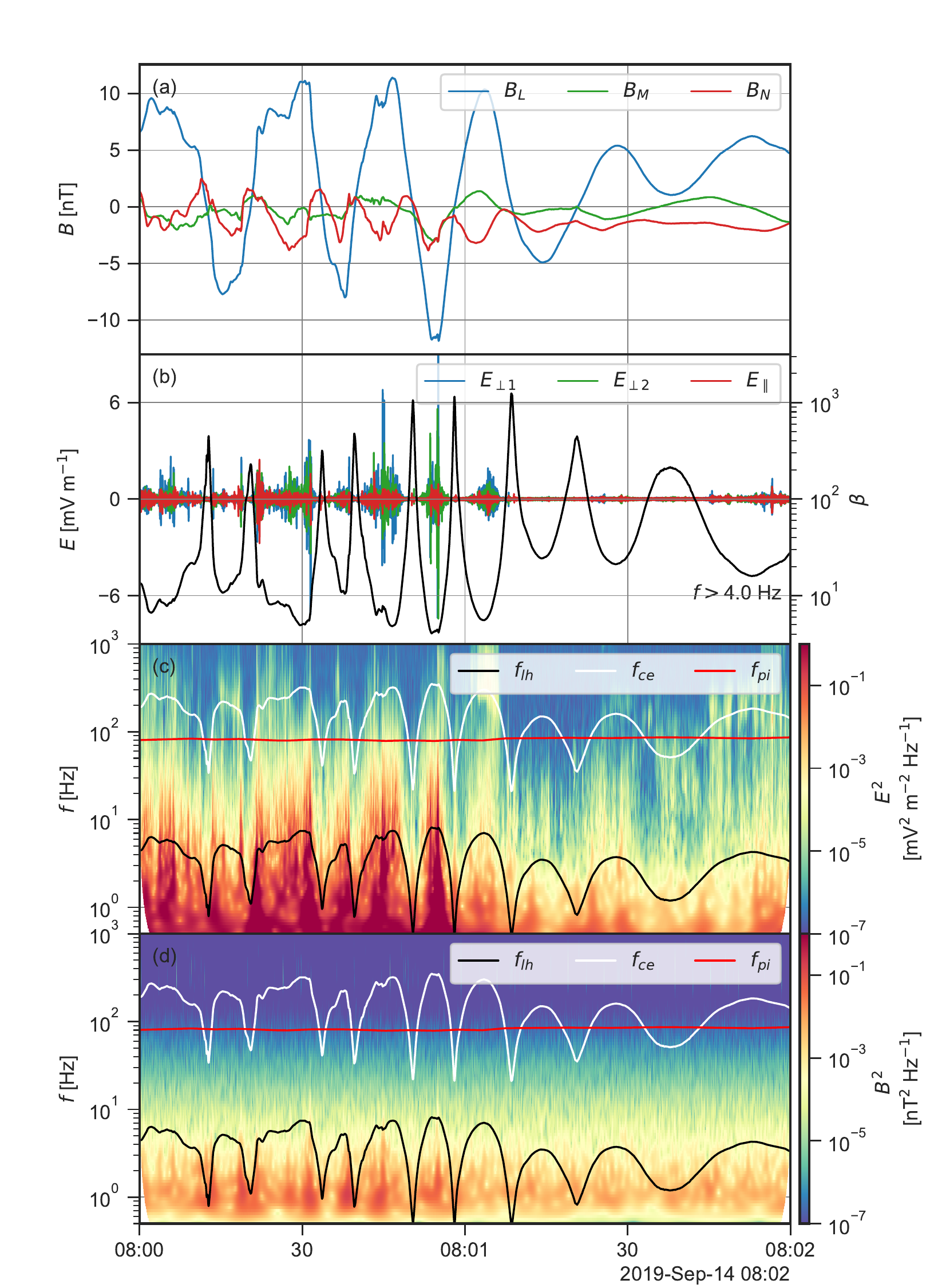}
	\caption{(a) Magnetic field in the global LMN coordinates system. (b) Electric field fluctuation above 4 Hz in field aligned coordinates together with the plasma $\beta$ (green). (c) Electric field spectrum and (d) magnetic field spectrum. The black, white and red lines in panels (c) and (d) show the lower hybrid frequency, the electron cyclotron frequency and the ion plasma frequency.}
	\label{fig:lhwaves}
\end{figure}

In addition to the low-frequency oscillations due to the flapping itself, higher frequency fluctuations of the electric field are observed at the current sheet boundaries. Figure~\ref{fig:lhwaves}b shows the electric field fluctuations $\delta \mathbf{E}$ above 4 Hz, together with the plasma parameter $\beta$ (green). We observe that the amplitude of the high-frequency oscillations of the electric field are almost zero at the local maxima of $\beta$ (current sheet center), while the amplitude of the high-frequency electric field peaks at the minima of $\beta$ (current sheet boundaries). This indicates that the high-frequency electric field waves grow at the current sheet boundaries and are suppressed within the current sheet which is consistent with \citeA{yoon_generalized_2002} and \citeA{daughton_electromagnetic_2003}.

We observe in Figure~\ref{fig:lhwaves}c that the frequency of the fluctuations of the electric field is close to the lower-hybrid frequency (shown by black line). Furthermore, we observe in Figure~\ref{fig:lhwaves}d that the magnetic field wave power is negligible ($\sim 10^{-4}$) compared with the electric field wave power, and hence the fluctuations are quasi-electrostatic. We identify these waves as lower hybrid drift waves (LHDWs). We note that a small magnetic component is expected even for the electrostatic LHDWs~\cite{norgren_lower_2012}. The LHDWs are seen only before 08:01:10, in the region where the current sheet thickness remains small ($h\sim 2~d_i$) and the density gradients observed in Figure~\ref{fig:ion-demagnetization}e are large. At later times, as the current sheet becomes thicker ($h > 5~d_i$) and the density gradients become smaller, the LHDWs vanish. This suggests that the LHDWs are driven by the density gradients arising from the thinning of the current sheet.

\section{Discussion}
\label{sec:discussion}
We found that the wave-like current sheet propagates along $\mathbf{M} = [0.87, 0.3, 0.38]$ GSM that is the average current direction. Also previous studies of flapping motions in the magnetotail found that the flapping motion is propagating along the current direction. In the magnetotail, the usual current direction is in the cross tail direction, and therefore the flapping waves propagate predominantly along the cross tail direction. In our case, the spacecraft are located on the dusk side of Earth ($\left [-6.10, 12.93, -0.14\right ]~R_E$), and there the average current direction is close to sunward and the flapping wave propagates sunward. This is consistent with the statistical surveys by \citeA{yushkov_current_2016} and \citeA{gao_distribution_2018}, who found that flapping motions far in the flank propagate toward the sun. The observed phase speed $\sim 250$ km s$^{-1}$ in the spacecraft frame is close to the ion bulk flow velocity, and thus the observed structure is nearly stationary in the ion frame of reference. Similar almost motionless flapping waves in the system of reference moving with the ion bulk flow were observed in PIC simulations \cite{sitnov_magnetic_2014}.

The observed wavelength of the flapping waves is ranging between $2.9 \times 10^3$ and $3.0 \times 10^4$ km (5.5 and 56 $d_i$). This is clearly much longer than the prediction given by \citeA{daughton_electromagnetic_2003} and elaborated by \citeA{sitnov_magnetic_2014} for a long-wavelength extension of the lower hybrid drift instability (LHDI) with $\lambda \sim 2\pi\sqrt{\rho_i \rho_e} \sim 1.63 d_i$. The shortest wavelength we observe is $\lambda \sim 5.72 d_i$ $(\sim 4.60 \rho_i)$ which is close to the flapping waves behind the dipolarization front in 3D PIC simulations \cite{sitnov_magnetic_2014}.

The average thickness of the observed current sheet is $6d_i$. The thickness varies during the event and reaches a minimum thickness $h = 1.8d_i = 1.45 \rho_i$. The observed geometry of the current sheet presents a wavy shape which resembles the one seen in simulations of thin $h\approx \rho_i$ current sheets \cite{lapenta_nonlinear_2002,sitnov_structure_2006}. The wavelength $\lambda$ of the observed wave-like current sheet scales with the thickness $h$ as $h\sim (0.18 \pm 0.03) \lambda$, or $kh\sim 1.15 \pm 0.21$, where $k=2\pi/\lambda$ is the wavenumber. This observed scaling of the wavelength of the wave-like current sheet with its thickness can be used in further development of the theory of the flapping motions.

This observed scaling can be used to characterize the instability which causes the observed wave-like structures. Linear stability analysis of an ion-scale Harris sheet, $h=\rho_i$, showed that the maximum growth rate of the kink mode corresponds to $kh\sim 1$ for $m_p/m_e=64$ \cite{daughton_unstable_1999}. Also, using a thin anisotropic current sheet model, \citeA{zelenyi_low_2009} showed that the maximum growth rate of the drift-kink instability belongs to the range $kh\in[0.8,~1.8]$ depending on the propagation angle. The observed scaling $kh\sim 1.15 \pm 0.21$ is close to the expected scaling for the drift-kink instability, which points at this instability as a possible cause of the observed wave-like structures.

Using the Ohm's law we have shown that a substantial non-ideal electric field arises $\mathbf{E} + \mathbf{V}_i \times \mathbf{B} \neq 0$ which is a strong evidence of the demagnetization of the ions and implies that the thickness of the current sheet is of the order of the ion scale. On the other hand, we observed that the non-ideal electric field is balanced by the Hall electric field, and hence that the electrons remain magnetized. Similar to the planar current sheet picture, the Hall electric field is normal to the kink-like current sheet. The observation of an electric field normal to the current sheet is consistent with the PIC simulation of super thin current sheet $(h < \rho_i)$ by \citeA{sitnov_structure_2006} who suggested that the electric field normal the current sheet is important for controlling the thickness of the thin current sheet.

Since the observed Hall electric field has a large component in the direction of propagation of the wave-like structure, and as discussed above is expected to be normal to the current sheet, it suggests that the observed current sheet is strongly tilted (tilt angle $\alpha > 45 ^\circ$, see Figure \ref{fig:current-sheet}). Hence, the local normal to the sheet is very different from the global $\mathbf{N}$ direction, i.e. the local normal is close to the global $\mathbf{M}$ direction (the direction of propagation of the wave-like structure) which we observe in Figure~\ref{fig:structure-velocity}b. This is consistent with the large tilt angle of the magnetotail current sheet reported by \citeA{shen_magnetic_2008} and resembles the strongly-tilted configuration ($\alpha \sim 90^\circ$) in Figure 16 of \citeA{sitnov_structure_2006}.

One possible mechanism to limit the thickness of the current sheet is through lower hybrid drift instability (LHDI). We observed such waves associated with strong gradients of density at the thinnest current sheet crossings. Similar strong gradients are found in the PIC simulation of thin current sheet \cite{sitnov_structure_2006}. The non-linear LHDI could modify the initial current sheet equilibrium and enhance the growth rate of the kink instability \cite{daughton_electromagnetic_2003}. The amplitude of the observed LHDWs changes with the plasma parameter $\beta$, consistent with damping in the high-beta region in the current sheet center \cite{bale_observation_2002}. The LHDI can be excited by the strong gradients, and will broaden these gradients \cite{winske_particle_1978}. Thus, our observations suggest that the LHDI can be important for limiting the thinning of the current sheet.

\section{Conclusion}
We have presented observations of a short-period (wave period $T\approx25$ s = 35 $f_{ci}^{-1}$) kink-like flapping of a thin current sheet in the flank of the Earth's magnetotail. The wave-like flapping structure propagates along the current direction.

We find that the observed current sheet thickness varies during the time we observe the flapping, and the minimum thickness is comparable to the ion inertial length. The wavenumber of the flapping oscillations scales with the current sheet thickness as $kh \sim 1.15 \pm 0.21$. Such a value of the wavenumber in terms of current sheet thickness belongs to the domain $kh \in [0.8, 1.8]$ of predicted maximum growth rate of the drift-kink instability, suggesting this instability can be responsible for the observed flapping.

The observed small value of the current sheet thickness is consistent with the observation of a decoupling between ions and electrons, which is related to the fact that the electrons are well magnetized but the ions are partially demagnetized. A large Hall electric field arises because of the small thickness of the current sheet. We observe that this Hall electric field is directed mainly in the direction of propagation of the wave-like structure which indicates that the current sheet is locally strongly tilted.

At the times when the current sheet thickness approaches the smallest ion-inertial-length scale we observed strong density gradients and associated lower hybrid drift waves (LHDWs) at the edges of the current sheet, which suggest that the LHDWs are generated in response to the thinning of the current sheet. Thus, the LHDWs can potentially provide a mechanism limiting the further thinning of the current sheet.

\acknowledgments
We thank the entire MMS team and instrument PIs for data access and support. All of the data used in this paper are publicly available from the MMS Science Data Center \url{https://lasp.colorado.edu/mms/sdc/}. We also wish to thank Victor Sergeev for valuable discussions. Data analysis was performed using the pyrfu analysis package available at \url{https://github.com/louis-richard/irfu-python}. The codes to reproduce the figures in this paper are available at \url{https://github.com/louis-richard/flapping}. This work is supported by the Swedish National Space Agency grant 139/18.

\bibliography{library}

\end{document}